\documentclass[twocolumn,showpacs,preprintnumbers,amsmath,amssymb]{revtex4}
\usepackage{graphicx}
\usepackage{epstopdf}
\usepackage{bm}

\usepackage{color}

\begin{document}

\title{Continued fraction analysis of dressed systems: application to periodically driven optical lattices}

\author{Thomas Zanon-Willette$^{1,2}$\footnote{E-mail address: thomas.zanon@upmc.fr \\}, Emeric de Clercq$^{3}$, Ennio Arimondo$^{4}$}
\affiliation{$^{1}$UPMC Univ. Paris 06, UMR 7092, LPMAA, 4 place Jussieu, case 76, 75005 Paris, France}
\affiliation{$^{2}$CNRS, UMR 7092, LPMAA, 4 place Jussieu, case 76, 75005 Paris, France}
\affiliation{$^{3}$LNE-SYRTE, Observatoire de Paris, CNRS, UPMC, 61 avenue de l'Observatoire, 75014 Paris, France}
\affiliation{$^{4}$Dipartimento di Fisica "E. Fermi", Universit\`a di Pisa, Lgo. B. Pontecorvo 3, 56122 Pisa, Italy}
\date{\today}



\begin{abstract}
Radio-frequency quantum engineering of spins is based on the dressing by  a non resonant electromagnetic field. Radio-frequency dressing occurs also for the motion of particles, electrons or ultracold atoms, within a periodic spatial potential. The dressing, producing a renormalisation and also a freeze of the system energy, is described by different approaches,  dressed atom, magnetic resonance semiclassical treatment, continued fraction solution of the Schr\"odinger equation.  A comparison between those solutions points out that the semiclassical treatment, to be denoted as the $S$-solution, represents the most convenient tool to evaluate the tunneling renormalization of ultracold atoms.

\end{abstract}

\maketitle

\section{Introduction}
The analysis of a system with few degrees of freedom, an electron or an atom, interacting with a large system, photons or phonons, relies often on a renormalization approach, where the parameters of the initial system are modified by the interaction. Examples of this approach are  the effective mass for the electron motion in a semiconductor, or the extensive renormalization in quantum electrodynamics. Another example is the dressed atom introduced by Cohen-Tannoudji in order to describe the modification of a two-level atomic magnetic response by an applied radio-frequency (rf) field in the absence of decoherence processes~\cite{CohenTannoudji1994}. For a non-resonant rf driving  at a high frequency Cohen-Tannoudji and Haroche~\cite{CohenTannoudji1966} derived a renormalization of  the atomic level splitting dependent on the amplitude of the rf field and described by a zero-th order ordinary Bessel function.  That modification producing a magnetic "freezing" of the two-level response, (i.e.  a nonmagnetic system), was examined for  atoms in~\cite{Haroche1970,Muskat1987,Esler2007,Chu2011}, for a Bose-Einstein condensate of chromium in~\cite{Beaufils2008}, for an artificial atom in~\cite{Tuorila2010}, and recently proposed for improving the precision of optical clocks~\cite{Zanon2012}.\\
\indent The same renormalization and freezing of the system properties under the application of a time-dependent modulation was  applied to a variety of processes, all  characterized by weak decoherence processes.  We mention here the dynamical localization describing the renormalized motion of a charged particle within a periodic potential under a time modulated force~\cite{Dunlap1986} and the coherent destruction of tunneling for a double well potential with a periodic driving, with a complete localisation of a wave packet in one well for specific values of the driving force~\cite{Grossman1991}.  For the motion of ultracold degenerated atomic gases within a shaken optical lattice,  the tunneling atomic evolution  is renormalized under the application of a time modulated force, as proposed in ~\cite{Grossman1991,Eckardt2005,Creffield2006,Kudo2009,Eckardt2010,Kudo2011} and tested experimentally in~\cite{Lignier2007,Kierig2008,Zenesini2009,Struck2011,Arimondo2012,Struck2012} within the framework of quantum simulation of solid state physics. The  renormalization of an optical lattice potential acting on cold atoms in a regime of classical diffusion and transport was investigated in~\cite{Wickenbrock2012}.
\\
\indent Target of the present work is to characterise the renormalization and freezing for a wide parameter range.  The starting point  is the  continued fraction solution for the two-level atomic magnetic response to an applied rf field, as typical of the nuclear magnetic resonance for a one-half spin. The regime of strong perturbation was investigated by several authors~\cite{AutlerTownes1955,Shirley1965,MorandTheobald1969,Stenholm1972,Swain1986}, in presence or absence of decoherence.   Their solution was expressed  in terms of infinite continued fractions. The present work investigates the renormalization process through  the continued fraction approach.  That treatment allows us to explore numerically the shaken-lattice renormalization for all parameter ranges, and in particular for explored experimental conditions. The numerical complexity of the continued-fraction solution, and its slow convergence in the regime of experimental interest, brought us to consider carefully the corrections to the zero-order Bessel function derived for the dressed-atom problem mainly through a semiclassical treatment~\cite{CohenTannoudji1973,Hannaford1973,Ahmad1974,Series1977}, and later recovered through renormalisation group techniques~\cite{Frasca2005}. On the basis of the analogy between the renormalization of the magnetic resonance energy and of the atomic tunneling in optical lattice, we focus our analysis from  the high frequency regime realized for a rf modulation at a very large frequency.
Then we explore the corrections when this limiting condition is not precisely satisfied. The low frequency regime covered by the treatments of refs.~\cite{Dunlap1986,Kayanuma2008,Kudo2009,Kudo2011} is not examined here.\\
\indent Sec. II presents different systems where the renormalization of the interaction strength has been investigated: magnetic resonance, motion in a periodic potential, tunnel coupling in a periodic potential. This section reports also the standard result of the renormalization process given by the zero-order Bessel function, valid under appropriate operating conditions. Sec. III reports the solution for the temporal evolution of the wavefunction in the magnetic resonance case. Sec. IV derives the renormalization through the  continued fraction approach, valid for all operating conditions, and also through a semiclassical treatment refening the Bessel-function result. Sec. V reports numerical results determining the limiting validity of the usual zero-order Bessel correction, and derive the renormalization for a large set of parameters. A conclusion completes our work.

\section{Dressed systems}
\subsection{Magnetic resonance}
\label{interaction}
{\it a) Semiclassical approach} For a spin-1/2 system interacting with a static magnetic field along the $z$ axis and driven by an oscillating rf field along the $x$ axis,  the semiclassical Hamiltonian $H_{sc}$ is
 \begin{equation}
 H_{sc}=\frac{\hbar \omega_0}{2} \sigma_z+\frac{\hbar \Omega}{2} cos(\omega t)\sigma_x.
 \label{semiclassical}
 \end{equation}
where $\sigma_{x,z}$ are the Pauli matrices and $\hbar \omega_0$ the energy splitting  between the magnetic levels and $\Omega$ the Rabi frequency   proportional to the rf field amplitude. \\
\indent By writing the atomic wavefunction $|\psi \rangle$
written   as a superposition of the $|\pm\rangle$ atomic eigenfunctions
\begin{equation}
|\psi(t)\rangle =\sum_{m=\pm}C_{\rm m}(t)|m\rangle,
\label{expansion}
\end{equation}
the Schr\"odinger equation  leads to the following temporal evolution for the $C_{\rm m}$ coefficients:
\begin{eqnarray}
i\dot{C}_{\rm +} &=  \frac{\omega_0}{2} C_{\rm +} +\frac{\Omega}{2} cos\left(\omega t\right)C_{\rm -},\nonumber \\
i\dot{C}_{\rm -} &=- \frac{\omega_0}{2} C_{\rm -} +\frac{\Omega}{2} cos\left(\omega t\right)C_{\rm +}.
\label{schrodinger}
\end{eqnarray}
\indent In the $\omega_0 \to 0$ limit, and for $C_{\rm +}(t=0)=1$ as initial condition, these equations have solution
\begin{equation}
|\psi(t)\rangle= cos\left[\frac{\Omega}{2\omega} sin\left(\omega t\right)\right]|+\rangle-isin\left[\frac{\Omega}{2\omega}  sin\left(\omega t\right)\right]|-\rangle.
\end{equation}
As in~\cite{Stenholm1972,Beaufils2008} the result of calculating the $\sigma_z$ time-averaged  mean value over $|\psi\rangle$  may be expressed through the following renormalized eigenvalues
of the $H_{\rm sc}$ Hamiltotian:
\begin{equation}
E^{\rm ren}_{\pm}=\pm \frac{\hbar \omega_0}{2} {\cal J}_{\rm 0}(\frac{\Omega}{\omega}).
\label{renormalization}
\end{equation}
The introduction of a renormalization coefficient $R$ defined by the ratio between renormalized and original eigenvalues leads to
\begin{equation}
{\cal R}= \frac{E^{\rm ren}_{\pm}}{\pm\frac{1}{2}\hbar \omega_0}={\cal J}_{\rm 0}(\frac{\Omega}{\omega})
\label{effectivemagnetic},
\end{equation}
depending on the zero-order ordinary Bessel function ${\cal J}_{\rm 0}$.  Thus the applied sinusoidal magnetic interaction renormalizes the atomic coupling to the static magnetic field, with a reduction by the factor ${\cal J}_0(\frac{\Omega}{\omega})$. The effective magnetic energy is frozen whenever $\Omega/\omega$ is a zero root of the ${\cal J}_0$ Bessel function, as observed in the experiments of refs.~\cite{Haroche1970,Muskat1987,Esler2007,Beaufils2008}. The magnetic resonance renormalization was explored by ref.~\cite{Chu2011} in the $\omega<\omega_0$ low-frequency regime, where  the present approach is not valid.\\

\indent{\it b) Quantized approach} Introducing a quantum description of the rf field, with operator $a^\dagger$ and $a$ for the creation
 and annihilation of one radiofrequency photon, the dressed-atom Hamiltonian $H_{da}$ of the above configuration is~\cite{Cohen1968}
  \begin{equation}
 H_{\rm da}=\hbar \omega a^\dagger a +\frac{\hbar \Omega}{\sqrt{2\bar n}}\left(a+a^\dagger\right)\sigma_x+\frac{\hbar \omega_0}{2} \sigma_z,
 \label{dressedHamiltonian}
 \end{equation}
where $\bar n$ represents the mean number of photons applied to the atoms.\\
\indent For the high-frequency case $\omega \gg \omega_0$ the last term in $H_{\rm da}$ may be neglected and its eigenstates easily determined. Then a perturbation treatment for  the $\sigma_z$ term of that Hamiltonian  leads to  Eq.~\eqref{renormalization} for describing the interaction with the static field~\cite{Cohen1968}. The previous renormalization result is obtained also through this approach.
 \indent Notice that the dressed atom approach, and also the semiclassical approach of~\cite{Series1977},  demonstrated that the ${\cal J}_0$ renormalization is valid for whatever spin value and equally spaced Zeeman levels.\\
\subsection{Dynamic Localization} Dynamic localization was  introduced by Dunlap and Kenkre~\cite{Dunlap1986}  for the motion of an electron on a discrete one-dimensional periodic lattice with spacing $d_L$ in the presence of an oscillating force. It is based on exact calculations for the particle motion. A single-particle basis useful for describing the electron tunneling among the discrete lattice sites is provided by the $j$-th Wannier function centered on the $j$ lattice site of the periodic potential~\cite{Ashcroft1976}.  In a given energy band the Hamiltonian for free motion on the periodic lattice is determined by tunneling matrix elements, which in general connect arbitrarily spaced lattice sites.
However, because the hopping amplitude decreases rapidly with the distance, the tunneling Hamiltonian
may be well approximated by including only the  $\hbar J$ tunneling energy hopping between neighboring lattice sites. Under this hypothesis, the Hamiltonian for the electron on the  linear lattice with an applied periodic force $F cos(\omega t)$  is~\cite{Dunlap1986,Raghavan1996}
\begin{eqnarray}
H_{dl} &=& \hbar J\sum_{m}\left(|j><j+1|+|j+1><j|\right) \nonumber \\
 &+&\hbar Kcos\left(\omega t\right) \sum_{j}j|j><j|.
\label{dynamiclocalization}
\end{eqnarray}  Here $\hbar K=Fd_{\rm L}$ is the time-modulated energy difference between neighboring lattice sites.  Dynamic localization entails a suppression of the particle transport with the particle position oscillating in time and returning periodically to its original value. It is associated  to particle motion on an infinite lattice and does not impose conditions on the frequency driving. Our focus based on the analogy with magnetic resonance is on the high frequency driving and on a lattice with a site finite number.\\
\indent  For the case of two lattice sites $(j=-1/2,1/2)$, introducing the Pauli operators, as $\sigma_z=|1/2><1/2|-|-1/2><-1/2|$ and so on, the above Hamiltonian becomes~\cite{Eckardt2005_2}
\begin{equation}
H_{dl} = \hbar J\sigma_x+\frac{\hbar \Omega}{2}cos\left(\omega t\right) \sigma_z,
\label{dynamiclocalization2}
\end{equation}
where we have introduced $\Omega=K$ in order to emphasize the equivalence of this Hamiltonian with that of Eq.~\eqref{semiclassical} apart a change of the quantization axes and  the $J \equiv \omega_0/2$ parameter correspondence. Therefore the dressed atom renormalization applies also to this system, leading to a renormalized tunneling rate,
\begin{equation}
J_{\rm eff}={\cal R}J.
\label{effective}
\end{equation}
Once again the system response is frozen whenever $\Omega/\omega=K/(\hbar \omega)$ is a root of the ${\cal J}_0$ Bessel function.The applied sinusoidal force produces a dynamic localization of the particle.\\
\indent  For a spin larger than one-half and more than two lattice sites, the Hamiltonian assumes a form equivalent to that of Eq.~\eqref{dynamiclocalization2},  except for the angular momentum. Thus the magnetic resonance analogy confirms the ${\cal R}$ renormalization also for an arbitrary number of lattice sites.\\
\subsection{Shaken optical lattice} In a 1D optical lattice ultracold atoms are confined within the potential minima created by a single laser standing wave with $d_{\rm L}$ spacing~\cite{Bloch2005,Morsch2006}.  The Hamiltonian for atomic motion on the periodic lattice is determined by tunneling matrix typically including only the  $J$  hopping between neighboring lattice sites.\\
\indent A periodic force $Fcos(\omega t)$ (to be referred as lattice shake) drives the atoms inside the optical lattice.  Using the $j$-th Wannier function  centered on the $j$ lattice site, the Hamiltonian of Eq.~\eqref{dynamiclocalization} describes also the motion of the ultracold atoms within the optical lattice, with $K=\Omega$  again the shaking  energy difference between neighboring sites of the linear chain. Therefore the dynamic localization and the renormalization of the previous Subsection applies also to the ultracold atoms shaken lattices~\cite{Grossman1991,Eckardt2005,Creffield2006,Kudo2009,Eckardt2010,Kudo2011}, as tested in several experiments~\cite{Lignier2007,Kierig2008,Zenesini2009,Struck2011,Arimondo2012,Struck2012}. The  parameters of the Hamiltonian of Eq.~\eqref{dynamiclocalization2} investigated in those experiments are reported in Table I. Notice that most experiments investigated the high-frequency regime, but large deviations from that regime also occurred.  Ref.~\cite{Creffield2010} pointed out the difficulties in the precise measurement  of the tunneling freeze from the ultracold atoms images. \\
\indent  \begin{table}
\caption{Values of the shaken-lattice experimental parameters, $J$  equivalent to $\omega_0/2$ , $\omega$,  $\Omega=K/\hbar$, measured in units of the  recoil frequency $\omega_{rec}/2 \pi$ of the investigated atom. The last column reports the $\omega_0/\omega$ ratio.}
\begin{tabular}{cccccc}
\hline
\hline
Ref.  & $J\equiv \omega_0/2$  & $\omega$ & $ \Omega$ &$\omega_0/\omega$  \\
\cite{Lignier2007}& 0.02-0.08 & 0.15-0.9 &0-6&  0.04-1.0  \\
\cite{Kierig2008} & 0.19 & 0.8-4 &0-1.2& 0.09-0.47 \\
\cite{Zenesini2009} &0.004 & 1.9 &0-3 &0.004  \\
\cite{Struck2011} & 0.002 & 0.9 &0-5& 0.004  \\
\cite{Struck2012} & 0.02 & 0.9 & 0-9& 0.04 \\

\hline
\hline
\end{tabular}
\end{table}
\section{Continued fraction approach}
\label{ContinuedFraction}
\begin{figure}
\centering
\resizebox{9.0cm}{!}{\includegraphics[angle=0]{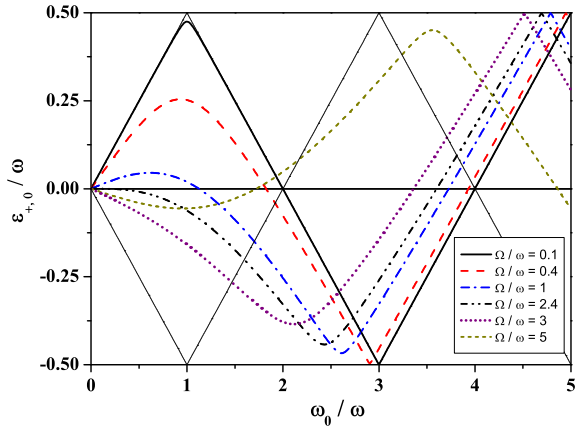}}
\caption{(Color online) Central Brillouin zone of the quasienergy   $\varepsilon_{+,0}$ vs $\omega_0$, both  measured in $\omega$ units,  for different values of $\Omega/\omega$, between 0 and 5. The quasienergy $\varepsilon_{-,0}$ is the opposite of $\varepsilon_{+,0}$. Quasienergies calculated by truncating the continued fraction to seven terms.  Freezing occurs  when the quasienergy is equal to zero at values different from $\omega_0=0$.  For $\Omega/\omega=2.4$ close to the ${\cal J}_0$ Bessel function first zero, owing to the quasienergy flatness at $\omega_0 \approx0$ a nearly perfect freezing is reached in a large range of low $\omega_0$ values.}
\label{diagramme-energie}
\end{figure}

If a single-particle Hamiltonian is
periodic in time, with  period $T$,
then the Floquet's theorem \citep{Floquet1883} states the existence
of a set of distinguished solutions $|\psi_n(t)\rangle$ to the time-dependent Schr\"odinger equation.
These Floquet states, analogous to the usual energy
eigenstates of time-independent Hamiltonian operators
\citep{Shirley1965,Sambe1973,Arimondo2012},
have the  form
\begin{equation}
	|\psi_n(t)\rangle = | u_n(t) \rangle
	\exp(-\rm i\varepsilon_n t/\hbar) \; .
\label{eq:FLS}
\end{equation}
with time periodic functions $|u_n(t)\rangle = |u_n(t+T)\rangle$. The quantum number $n$
specifies the state. The quantities $\varepsilon_n$
 are denoted quasienergies. By inserting Eq.~\eqref{eq:FLS}
into the Schr\"odinger equation governed
by the $H_{sc}$ Hamiltonian, we deduce
\begin{equation}
	\left( H_{sc} - \rm i\hbar\frac{\partial}{\partial t} \right)
	| u_n (t) \rangle
	= \varepsilon_n | u_n (t) \rangle  \; ,
\label{eq:EVE}
\end{equation}	
to be regarded as an eigenvalue equation\ for
the Floquet quasienergies. The set of Floquet functions  is
complete in the Hilbert space on which acts the Hamiltonian. Hence,
any solution $|\psi(t)\rangle$ to the Schr\"odinger
equation admits an expansion  in the $|u_n(t)\rangle$ basis.\\
\indent If  $| u_n (t) \rangle $ be a solution to the
eigenvalue Eq.~\eqref{eq:EVE} with quasienergy $\varepsilon_n$,  then $| u_n (t) \rangle e^{\rm i m \omega t} $ also is
a $T$-periodic solution, with quasienergy $\varepsilon_n + m\hbar\omega$  $m$ being an arbitrary
integer, where $\omega = 2\pi/T$.  Therefore
the quasienergy of a Floquet state is determined only up to an integer multiple of the  $\hbar\omega$
photon energy. In accordance with the solid-state physics terminology, the quasienergy spectrum is said to consist of an infinite
set of identical Brillouin zones of width $\hbar\omega$, covering the
entire energy axis, each state placing one of its quasienergies
in each zone.\\
\indent The quasienergies may be determined by diagonalization of the Hamiltonian expressed in the Fourier space, as in~\cite{Shirley1965}, or equivalently diagonalizing the dressed-atom Hamiltonian as in~\cite{Cohen1968}.  We will make use of the continued fraction solution of refs.~\cite{AutlerTownes1955,Stenholm1972}.  We apply the Fourier expansion to   the $C_{\pm}$ coefficients  of Eq.~\eqref{expansion}
 \begin{equation}
\begin{split}
C_{-}(t)=e^{i\lambda t}\sum_{n=-\infty}^{n=+\infty}A_{n}e^{-in\omega
t},\hspace{0.5cm}C_{+}(t)=e^{i\lambda t}\sum_{n=-\infty}^{n=+\infty}B_{n}e^{-in\omega
t}.
\end{split}
\end{equation}
Substituting these expansions into Eq.~\eqref{schrodinger} and equating the same  order Fourier components, one obtains
\begin{subequations}\label{}
\begin{gather}
\begin{align}
\left(\lambda-\frac{\omega_{0}}{2}-n\omega\right)A_{n}&=-\frac{\Omega}{4} B_{n-1}-\frac{\Omega}{4}  B_{n+1},\\
\left(\lambda+\frac{\omega_{0}}{2}-n\omega\right)B_{n}&=-\frac{\Omega}{4} A_{n-1}-\frac{\Omega}{4}  A_{n+1}.
\end{align}
\end{gather}
\end{subequations}
These equations can be separated into a  first set with all even $A'$s and odd $B'$s being zero
\begin{subequations}\label{L1}
\begin{gather}
\begin{align}
\left({\tilde \lambda}_{+}-\frac{4\omega_{0}}{\Omega}-l\frac{4\omega}{\Omega}\right)A_{l}&=-B_{l-1}-B_{l+1},\\
\left({\tilde \lambda}_{+}-k\frac{4\omega}{\Omega}\right)B_{k}&=-A_{k-1}-A_{k+1}.
\end{align}
\end{gather}
\label{lambdaplus}
\end{subequations}
\noindent and into a second one with all odd $A'$s and even $B'$s being zero
\begin{subequations}\label{L2}
\begin{gather}
\begin{align}
({\tilde \lambda}_{-}-k\frac{4\omega}{\Omega})A_{k}&=-B_{k-1}-B_{k+1},\\
({\tilde \lambda}_{-}+\frac{4\omega_{0}}{\Omega}-l\frac{4\omega}{\Omega})B_{l}&=-A_{l-1}-A_{l+1},
\end{align}
\end{gather}
\label{lambdaminus}
\end{subequations}
Here $k$ even,  $l$ odd,  and we introduced
\begin{eqnarray}
{\tilde \lambda}_{+}&=\frac{4}{\Omega}\left(\lambda+\frac{\omega_{0}}{2}\right), \\ \nonumber
 {\tilde \lambda}_{-}&=\frac{4}{\Omega}\left(\lambda-\frac{\omega_{0}}{2}\right).
 \end{eqnarray}
 \indent Eqs.~\eqref{lambdaplus} and ~\eqref{lambdaminus} are independent and a complete solution is obtained by adding the solutions of those equations. Eqs.~\eqref{lambdaminus} may be rewritten as
\begin{equation}
\label{xrecurrence}
x_{j}=-\frac{x_{j-1}+x_{j+1}}{D_{j}}
\end{equation}
by imposing  $x_{j}\equiv A_{j}$ for even $j$, $x_{j}\equiv B_{j}$ for odd $j$, with
\begin{subequations}\label{}
\begin{gather}
\begin{align}
D_{j}&\equiv {\tilde \lambda}_{-}+4\omega_{0}/\Omega-j4\omega/\Omega\hspace{1cm}\text{for odd j },\\
D_{j}&\equiv {\tilde  \lambda}_{-}-j4\omega/\Omega\hspace{1cm}\text{for even j}.
\end{align}
\end{gather}
\end{subequations}
\indent The recurrence Eq.~\eqref{xrecurrence} has a continued fraction solution~\cite{StenholmLamb1969,Stenholm1972}, with expression for $j>0$
\begin{equation}\label{}
\frac{x_{j}}{x_{j-1}}=-\frac{1}{D_{j}-\frac{1}{D_{j+1}-\frac{1}{D_{j+2}-\frac{1}{...}}}}\label{1}
\end{equation}
and similar expression for a negative $j$. By replacing $x_1$ and $x_{-1}$  into  Eq.~\eqref{xrecurrence} for $j=0$, we obtain  the continued fraction solutions  for ${\tilde \lambda}_+$ and ${\tilde \lambda}_-$, with
\begin{eqnarray}
{\tilde  \lambda}_{-}&=&\frac{1}{{\tilde  \lambda}_{-}+\frac{4\omega_{0}}{\Omega}\left(1-\frac{\omega}{\omega_{0}}\right)-\frac{1}{{\tilde  \lambda}_{-}-\frac{8\omega}{\Omega}-\frac{1}{{\tilde  \lambda}_{-}+\frac{4\omega_{0}}{\Omega}\left(1-3\frac{4\omega}{\Omega}\right)-\frac{1}{...}}}}\\ \nonumber
&+&\frac{1}{{\tilde  \lambda}_{-}+\frac{4\omega_{0}}{\Omega}\left(1+\frac{\omega}{\omega_{0}}\right)-\frac{1}{{\tilde  \lambda}_{-}+\frac{8\omega}{\Omega}-\frac{1}{{\tilde  \lambda}_{-}+\frac{4\omega_{0}}{\Omega}\left(1+3\frac{4\omega}{\Omega}\right)-\frac{1}{...}}}},
\label{Autler-Townes-equation}
\end{eqnarray}
and
\begin{equation}
{\tilde \lambda}_+=-{\tilde \lambda}_-.
\end{equation}
All Floquet quasienergies are given by
\begin{equation}
\varepsilon_{\pm,n}=-\hbar\lambda=\hbar\left(\pm\frac{\omega_0}{2}-\frac{\Omega}{4}{\tilde \lambda}_{\pm}\right)+n\hbar\omega
\end{equation}
The continued fraction  solution allows to determine numerically the Floquet quasienergies with the required accuracy. Fig.~\ref{diagramme-energie} reports the quasienergies within one Brillouin zone vs $\omega_0$ for  different values of the $\Omega/\omega$ parameter. Those energy diagrams may be applied to analyse either magnetic resonance or dynamical localisation or shaken optical lattices. The zero crossing of the energy represent magic values where the effective magnetic energy or quantum tunneling are frozen at values different from $\omega_0=J=0$.
 \begin{figure}
\centering%
\resizebox{7.0cm}{!}{\includegraphics[angle=0]{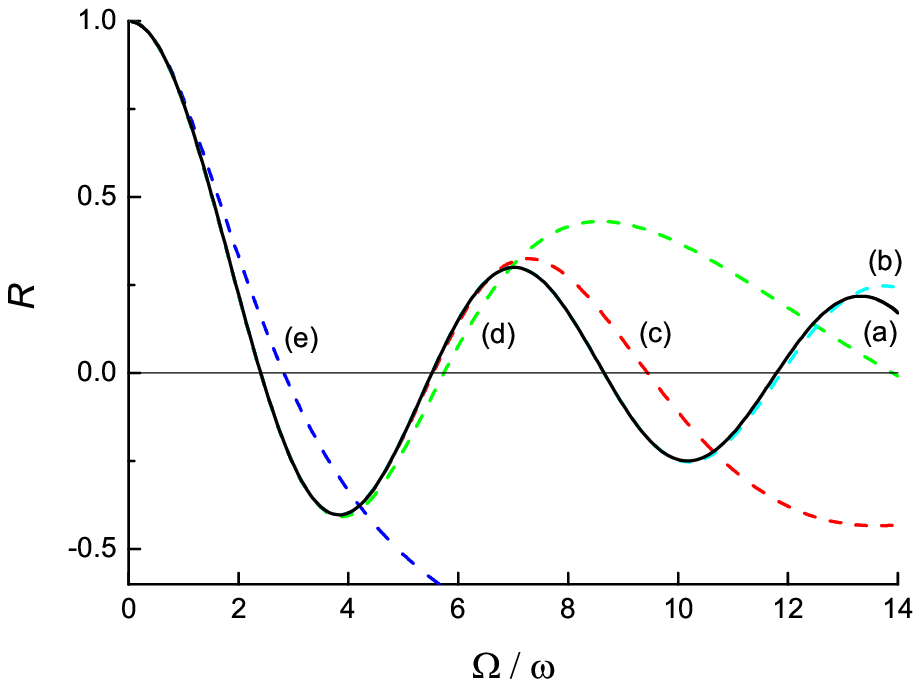}}
\resizebox{7.0cm}{!}{\includegraphics[angle=0]{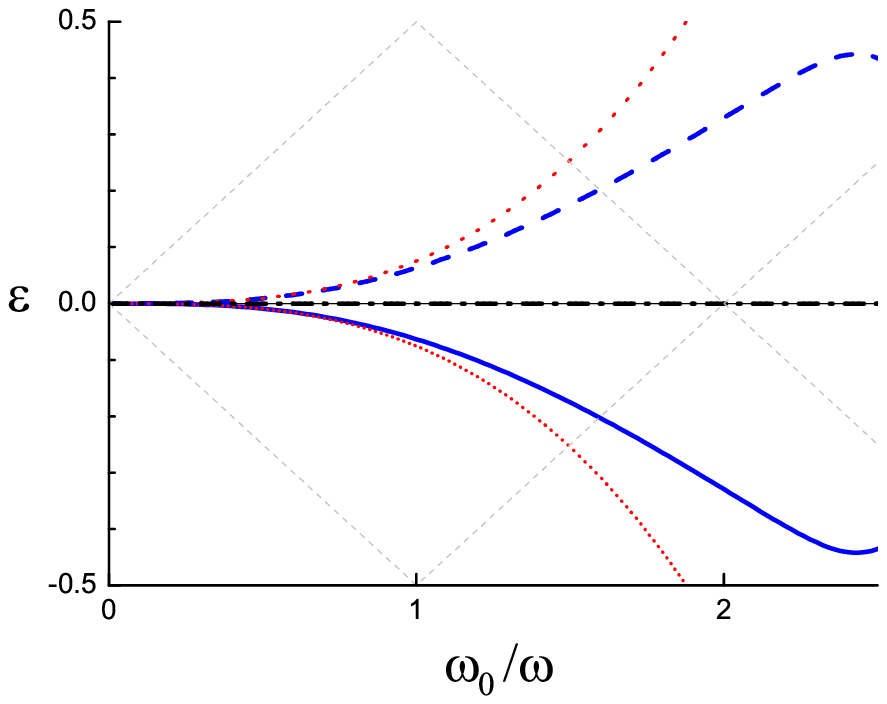}}
\caption{(Color online) On the top ${\cal R}$ vs $\Omega/\omega$ calculated at $\omega_0/\omega=0.1$ on the basis of different approximations: (a) from the continued fraction  containing terms up to 9 terms, the ${\cal J}_0$ solution of Eq.~\eqref{renormalization} and the $S$-corrected one of Eq.~\eqref{approxgeff}; (b), (c), (d) and (e)   from the continued fraction limited to seven four, three and one terms, respectively.  On the bottom quasienergies $\varepsilon_{+,0}$ (continuous lines) and  $\varepsilon_{-,0}$ (dashed lines) vs $\omega_0$ at $\Omega/\omega=2.405$. Thicker blue lines calculated on the basis of the four terms continued-fraction; thinner red lines on the $S$-corrected solution. The central black horizontal dot-dashed line based on Eq.~\eqref{zeroBessel}. Diagonal lines for the $\omega_0 \to 0$ quasienergies.}
\label{Facteur-de-Lande}
\end{figure}
\section{Renormalization vs  $\omega_0$ value}
The energy renormalization, to be investigated on the basis of different theoretical approaches will be concentrated on magnetic resonance case, but the analysis of Sec. II has demonstrated that the substitution $J=\omega_0/2$ allows to apply our results also to the shaken optical lattices.\\
\indent {\it a) ${\cal J}_0$ solution} The two-level energy splitting at $\omega\gg\omega_0$ derived in Eq.~\eqref{renormalization} by the magnetic resonance treatment  leads to the ${\cal R}$ renormalization given by ${\cal J}_0$ Bessel function of Eq.~\eqref{effectivemagnetic}.  The ${\cal J}_0$ renormalization approximation corresponds to the following quasienergies:
\begin{equation}
\frac{\varepsilon_{\pm,n}}{\hbar}=\pm\frac{\omega_0}{2}{\cal J}_{0}\left(\frac{\Omega}{\omega}\right)+n\omega.
\label{zeroBessel}
\end{equation}
This solution predicts a freezing for whatever $\omega_0$ at the $\Omega/\omega$ values corresponding to the zeros of the ${\cal J}_0$ Bessel function, but its  validity is limited to $\omega_0 \approx 0$. \\
\indent{\it b) $S$-corrected solution}  On the basis of a magnetic resonance semiclassical treatment, refs.~\cite{Hannaford1973,Series1977} derived an $\omega_0$-dependent correction to the ${\cal J}_0$ renormalization.  That correction leads to the following quasienergies and renormalization:
\begin{eqnarray}
\frac{\varepsilon_{\pm,n}}{\hbar}&=&\pm\frac{\omega_0}{2}\left[{\cal} J_{0}\left(\frac{\Omega}{\omega}\right)-\left(\frac{\omega_0}{\omega}\right)^2S\left(\frac{\Omega}{\omega}\right)\right]+n\omega, \\
{\cal R}&=&{\cal J}_{0}\left(\frac{\Omega}{\omega}\right)-\left(\frac{\omega_0}{\omega}\right)^2S\left(\frac{\Omega}{\omega}\right).
\label{approxgeff}
\end{eqnarray}
Here $S(x)$ is a product of ${\cal J}_n$ ordinary Bessel functions well approximated by the following expression~\cite{Ahmad1974}:
\begin{equation}
S(x)=\frac{16}{2025x^4}\left[\alpha(x){\cal J}_2(x)+\beta(x){\cal J}_4(x)-\gamma(x){\cal J}_6(x)\right];\\
\label{S-function}
\end{equation}
where $\alpha(x)~=~75\left(5-x^2/4\right)x^{2}$, $\beta(x)=6\left(408-74x^{2}-23x^{4}/16\right)$ and $\gamma(x)=145x^{2}\left(3-x^{2}/2\right)/49$.  Within the following Section this solution will be used  for calculations around the first and second zeros of the zero-th order Bessel function, and  the validity limits for the $\omega_0/\omega$ application range will be discussed there.  \\
\indent {\it c) Continued fraction} A general approach to derive the ${\cal R}$ renormalization  coefficient is based on the Floquet quasienergies derived in previous Section, leading to
\begin{equation}
{\cal R}=\frac{\varepsilon_{+,0}}{\hbar \omega_0/2}.
\label{Rtoquasienergies}
\end{equation}
This equation allows a numerical determination of ${\cal R}$ without restricting to the low $\omega_0$ values where the ${\cal J}_0$ approximation (without or with the $S$ function)  is valid. The quasienergy can be derived either from the continued fraction solution of Eq.~(20), truncated to a finite number $j$ of terms, or from a  diagonalization of the system of Eqs.~\eqref{xrecurrence} truncated to a finite number of equations with $2j+1$ terms.  The Hamiltonian diagonalization approach was applied in ref.~\cite{Chu2011} for the renormalisation calculations in both the low-frequency and high-frequency regimes.
\begin{figure}[!t]
\centering%
\hfill\resizebox{8.0cm}{!}{\includegraphics[angle=0]{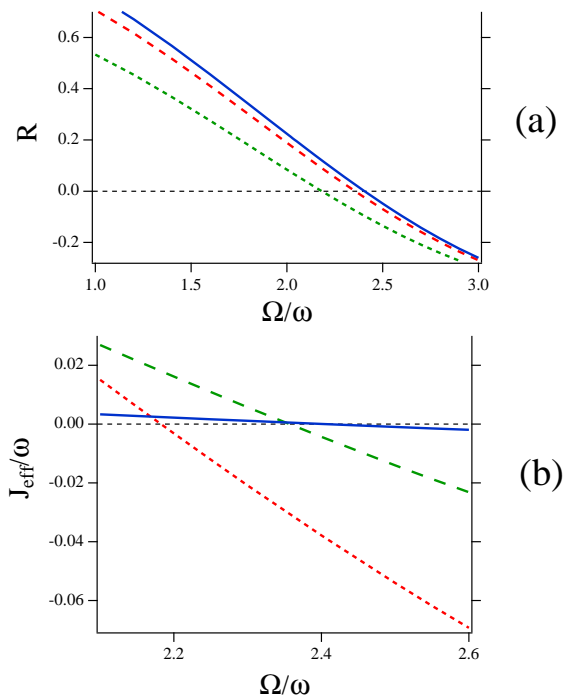}}
\caption{(Color online) In (a) renormalization coefficient ${\cal R}$ and  in (b) effective tunneling ${\cal J}_{eff}$ vs $\Omega/\omega$ ratio for different values of $J/\omega$.  Continuous blue lines for $J/\omega=0.02$, dashed green lines for $J/\omega=0.2$ and dotted red lines for $J/\omega=0.4$. }
\label{shakenrenormalization}
\end{figure}
\begin{figure}[!t]
\centering%
\resizebox{9.0cm}{!}{\includegraphics[angle=0]{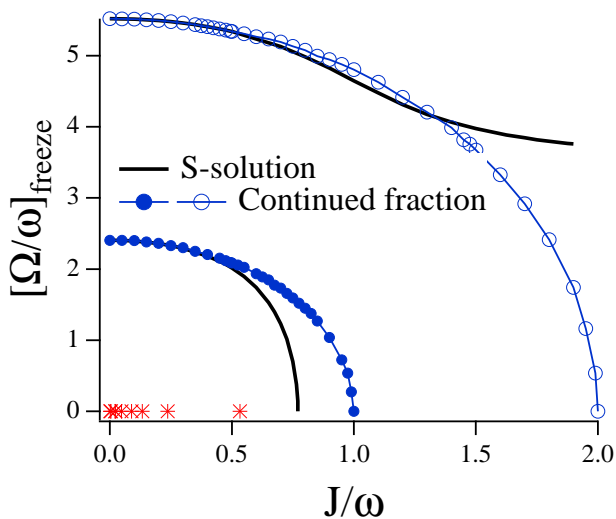}}
\caption{(Color online) $\left[\Omega/\omega\right]_{freeze}$ values required to freeze the tunneling coefficient ($J_{eff}=0$) as a function of the $J/\omega$ unperturbed tunneling coefficient. Continuous lines from the $S$ corrected expression and dots from the continued fraction solution. Closed blue dots for the first zero crossing of the eigenenergies and open blue ones for the second zero crossing. The stars on the horizontal axis denotes the $J/\omega$ values explored in the shaken optical lattice experiments.}
\label{shakenfreezing}
\end{figure}

\section{Numerical results}
The ${\cal R}$ renormalization  coefficient is a complex function of the system parameters  $\omega_0/\omega$ and $\Omega/\omega$, and the
previous Section approaches may be used  for numerical analyses at different parameter values. Fig 2(a) reports the $R$ results at $
\omega_0/\omega=2J/\omega=0.1$ vs $\Omega/\omega$, obtained using Eq.~\eqref{Rtoquasienergies} linking that coefficient to the Floquet quasienergies. For the quasienergies  determined from the continued fraction solution, and also from the diagonalization of the system of Eqs.~\eqref{xrecurrence}, the results of Fig. 2(a) shows a slow convergence at low $\omega_0$ values, as already pointed out by Autler and Townes~\cite{AutlerTownes1955},  the number of required terms in the continued fraction depending on the  $\Omega/\omega$ value.  In the $\omega_0,J \to 0$ limit ${\cal R}$ is well approximated by the ${\cal J}_0(\Omega/\omega)$ function, the $S$ correction vanishing there.\\
\indent  For the quasienergies dependence on $\omega_0$ at $\Omega/\omega=2.405$, Fig.~2(b) compares the continued fraction solution to the ${\cal J}_0$ solution and the $S$-corrected one. The ${\cal J}_0$ solution leads to a horizontal line close to $x$-axis because $\Omega/\omega$ corresponds to the Bessel first zero, indicating that it approximates the quasienergies only for $\omega_0\approx0$.  On the contrary the $S$-corrected solution  approximates well the quasienergies for a large range of parameters, at least for $\omega_0/\omega$  up to 0.5 corresponding to $J/\omega$ up to 1.\\
\indent For the shaken optical lattice experiments where  the condition $\omega_0=2J \ll \omega$ is satisfied, as in most cases, the ${\cal J}_0$ solution is well appropriate for ${\cal R}$. At larger $J$ values, the $S$ function correction to ${\cal J}_0$ can be used for the full range of the parameters explored so far in experiments. Fig.~\ref{shakenrenormalization} reports a $S$-correction based analysis of the shaken-lattice renormalization at increasing values of $J/\omega$.
For $J/\omega=0.4$ the  correction  to ${\cal R}$ shown in Fig.~\ref{shakenrenormalization}(a) is ten percent smaller than the $J/\omega\approx 0$ value, but  becomes larger increasing $J$.  Because the most important quantity is the tunneling coefficient itself, Fig.~\ref{shakenrenormalization}(b) shows the $J_{eff}\approx 0$ dependence  on $\Omega/\omega$  at increasing values of $J/\omega$. Notice that increasing $J/\omega$ the $J_{eff}=0$ freezing configuration is reached at an $\Omega/\omega$  value lower than the Bessel first zero.\\
\indent  Fig.~\ref{shakenfreezing} shows the $\left[\Omega/\omega\right]_{freeze}$ values required to produce a tunneling freeze for a given $J/\omega$ initial value. We plot the values associated to the first and second zero-crossing of the eigenenergies, corresponding to the first and second zero of the Bessel function within the ${\cal J}_0$ solution.  A  comparison between the $S$-corrected solution and the continued fraction solution is presented, confirming that for most shaken-lattice experiments performed so far, the $S$-corrected solution provides a simple and precise determination of the modified tunneling parameter. For a larger range of parameters the continued fraction solution should be used.  The data points at $\Omega/\omega \to 0$ correspond to the quasienergy crossings in absence of rf drive and don't have a physical meaning. Notice that freezing can be produced also applying $\omega$ values lower than $J$, a  regime was  not yet examined in the experiments. It may be noticed that the general dependence of the freezing value of Fig.~\ref{shakenfreezing} is similar, although not identical to the Bloch-Siegert shift dependence investigated in~\cite{CohenTannoudji1994,CohenTannoudji1966,AutlerTownes1955,Shirley1965,Stenholm1972,Swain1986,Series1977,CohenTannoudji1973,Hannaford1973,Ahmad1974}. In fact for an applied oscillating field, as in the present magnetic resonance configuration, all the crossings and anticrossings of the energy levels are shifted towards lower $\omega_0$ values by increasing $\Omega$~\cite{Cohen1968}. The Bloch-Siegert shift of the magnetic resonance is associated to the position of the energy anticrossings, while the freezing point is associated to the zero crossing of the eigenenergies.

\section{Conclusions}
We have examined the energy renormalisation of a two-level system usually associated to the dressed atom approach, but also derivable from a semiclassical analysis of the magnetic resonance. The $S$-solution derived in ref.~\cite{Zanon2012} for the Zeeman freezing of optical clocks is here applied to the optical lattice experiments. Within that framework the standard zero-Bessel dependence on the amplitude of the electromagnetic field amplitude, valid only at zero magnetic field, was extended to derive a general dependence on the magnetic field amplitude. That result is important for the main target of the present work, to use the magnetic resonance results in order to perform an accurate analysis of the renormalisation occurring for the atomic quantum tunneling between the minima of an optical lattice in the shaken lattice experiments. The magnetic resonance correction to the energy renormalisation allows us to derive a very  general formula for the tunneling renormalisation in shaken optical lattices. The conditions for the complete cancellation of the tunneling rate are functions of the tunneling energy without shaking and of the  modulation frequency. The precise determination of the tunneling under different driving conditions will lead to a better control in the quantum simulation experiments based on optical lattices.

\section{Acknowledgments}
The authors are grateful to Donatella Ciampini and Mich\`ele Glass--Maujean for a careful reading of the manuscript and suggestions.



\end{document}